# Single-Particle Dynamics in a Nonlinear Accelerator Lattice: Attaining a Large Tune Spread with Octupoles in IOTA

Sergey A. Antipov, University of Chicago, Chicago, Illinois
Sergei Nagaitsev, Alexander Valishev, Fermilab, Batavia, Illinois

Abstract

Fermilab is constructing the Integrable Optics Test Accelerator (IOTA) as the centerpiece of the Accelerator R&D Program towards high-intensity circular machines. One of the factors limiting the beam intensity in present circular accelerators is collective instabilities, which can be suppressed by a spread of betatron frequencies (tunes) through the Landau damping mechanism or by an external damper, if the instability is slow enough. The spread is usually created by octupole magnets, which introduce the tune dependence on the amplitude and, in some cases, by a chromatic spread (tune dependence on particle's momentum). The introduction of octupoles usually has both beneficial (improved Landau damping) and harmful properties, such as a resonant behavior and a reduction of the dynamic aperture. One of the research goals at the IOTA ring is to achieve a large betatron tune spread, while retaining a large dynamic aperture, using conventional octupole magnets in a special but realistic accelerator configuration. The configuration, although not integrable by design, approximates an autonomous 2D Hamiltonian system. In this paper, we present results of computer simulations of an electron beam in the IOTA by particle tracking and the Frequency Map Analysis. The results show that the ring's octupole magnets can be configured to provide a betatron tune shift of 0.08 (for particles at large amplitudes) with the dynamical aperture of over 20 beam sigma for a 150-MeV electron beam. The influence of the synchrotron motion, lattice errors, and magnet imperfections is insignificant for the parameters and levels of tolerances set by the design of the ring. The described octupole insert could be beneficial for enhancing Landau damping in high intensity machines.

**Introduction**

Modern circular accelerators are designed with a linear transverse focusing optics. In an ideal case with linear optics, the single particle motion is fully integrable, that is, there are as many independent conserved dynamic quantities (the Courant-Snyder invariants), as the number of degrees of freedom. For an integrable system the motion is always "solvable" and regular with no chaotic behavior, independent of initial conditions. In reality, focusing lattices are far from being perfectly linear. If a system is linear by design, small nonlinear perturbations break the integrability and may lead to a resonant behavior, chaotic and unbounded motion. In accelerators such perturbations arise from high-order magnet errors and imperfections or are introduced on purpose (i.e. sextupole magnets for chromaticity correction), leading to a dependence of a single-particle stability on initial conditions [1, 2].

There are many examples of nonlinear integrable dynamical systems. One such example of an integrable nonlinear system is the Kepler problem, where $U \propto 1/r$. The motion is regular (not necessarily bounded), and for bounded orbits the oscillation frequency strongly depends on the amplitude: $\epsilon \propto a^{-3/2}$. Obviously, such a potential is of limited use in particles accelerators because of its singularity at the origin.

Generally, there has not been a lot of success in funding integrable nonlinear focusing lattices for circular accelerators. Nonetheless, in the late 1960's McMillan discovered a one-dimensional nonlinear mapping – a thin kick of a special kind in an otherwise linear lattice leading to a conservation of an integral of motion, quadratic in particle's coordinate and momentum. This solution was later generalized to 2D-case of an uncoupled symmetric lattice [3]. The required nonlinear kick can be created by, for example, an electron lens [4] but difficult to implement with magnets.

Recently, in 2010 a new nonlinear accelerator focusing system, which can be implemented with magnets, was proposed [5]. It consists of a linear focusing lattice with a transfer matrix of a thin axially symmetric lens, followed by a nonlinear magnetic field region, which can be created by a special magnet, and, in 2D, yields a conservation of the Hamiltonian and the second invariant of motion, quadratic in momenta. In contrast to the traditional linear focusing, it provides large a amplitude-dependent tune spread, which is beneficial for Landau damping of collective instabilities, with single particle motion being unconditionally stable and non-chaotic at all amplitudes. This concept will be experimentally tested at the Integrable Optics Test Accelerator (IOTA) at Fermilab [6].

The potential in [5] has two points of singularity, making the design, manufacturing, and testing of the special magnet rather challenging [7]. As a consequence, the first stage of the experiment will use conventional octupole magnets to create a corresponding lowest-order nonlinear term in the multipole expansion of the potential. In theory, this approach allows to achieve single particle dynamics with one integral of motion (for two transverse degrees of freedom), which is not sufficient for integrability. Even a single integral of motion can significantly improve particle dynamics in terms of achievable tune shifts and dynamical aperture. For instance, it proved to be beneficial for achieving a record-high beam-beam tune shift with round colliding beams at VEPP-2000 (BINP, Russia) [8].

Since Ref. [5] considered only transverse motion, it was important to make sure that the introduction of the 3$^{rd}$, longitudinal, degree of freedom does not destroy the integrability. Earlier works on simulations of dynamics in IOTA [6], [9] did not consider the pure-octupole case, and treated the ring as a linear transfer matrix or disregarded the longitudinal motion. A thorough study of dynamics in a realistic machine lattice in the presence of coupling, chromaticity, magnet imperfections, and parasitic nonlinearities was required to estimate the influence of these effects. In this paper we present the results of 3D particle tracking simulations of single particle dynamics in a nonlinear optics experiment at IOTA.

**Octupole experimental setup in IOTA**

The IOTA is a relatively small storage ring with the circumference of 40 m, designed to circulate either electrons or protons of the same momentum, ~150 MeV/c. In the electron mode, it will operate with short bunches of 150 MeV electrons injected from a superconducting linac [10]. The ring has a flexible linear optics, allowing for the installation of one or two nonlinear magnets with the transfer matrix of the rest of the ring outside of those magnets being an equivalent of a thin axially symmetric lens. The ring's focusing optics has a two-fold symmetry; its layout and optics functions are shown in Fig. 1, and main parameters of the machine are summarized in Table 1. A detailed description of the ring can be found in [9].

At the first stage of its experimental program the IOTA will study beam dynamics in a nonlinear lattice with one integral of motion, proposed in [5]. The nonlinearity will be created by a set of octupole magnets, placed in one of the 1.8 m long drifts with equal beta-functions and zero dispersion (Fig. 1). Their strength will vary inversely proportional to $\beta^3$, creating a potential:

$$V(x, y; s) = \frac{\mathrm{r}}{\mathrm{s}(s)^3}\left(\frac{x^4}{4} + \frac{y^4}{4} - \frac{3x^2 y^2}{2}\right) \qquad (1)$$

In the normalized coordinates $U = \mathrm{s}(s)V(x, y; s)$, and the resulting Hamiltonian $H$ (neglecting the longitudinal motion) is

$$H = H_0 + U = \frac{1}{2}\left(P_x^2 + P_y^2 + x_N^2 + y_N^2\right) + \mathrm{r}\left(\frac{x_N^4}{4} + \frac{y_N^4}{4} - \frac{3x_N^2 y_N^2}{2}\right), \qquad (2)$$

where $x_N$, $y_N$, $P_x$, and $P_y$ are the normalized transverse coordinates and momenta. The Hamiltonian $H$ is independent of the longitudinal position $s$ and, thus, is an integral of motion in the octupole channel. Because of the lack of the second integral, this Hamiltonian system is non-integrable. The betatron phase advance in the channel is equal to 2 ×0.3 both in vertical and horizontal planes. The rest of the ring has a linear transfer matrix of a thin axially symmetric lens (Fig. 2) with the phase advance of 2 ×5 and, therefore, preserves the integral of motion to first order. In reality, lattice errors, imperfections, and parasitic nonlinearities in the ring elements will distort the system. But for small enough perturbations, $H$ may remain an approximate integral of motion and the motion will remain close to unperturbed.

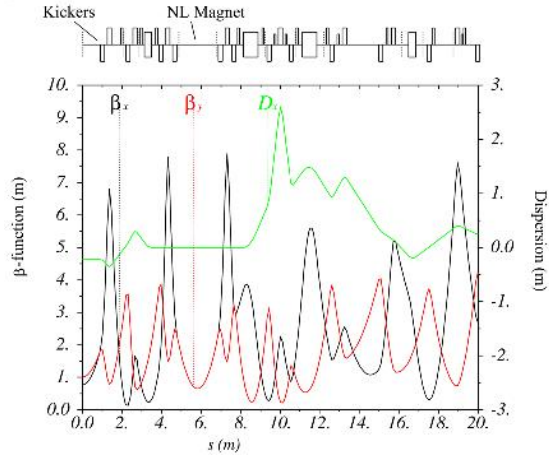

Figure 1: The beta functions (horizontal – black line, vertical – red) are equal and the horizontal dispersion (green) is zero in the nonlinear insert. A half of the IOTA ring for the setup of one nonlinear insert is depicted; ring elements and their positions are shown on the top.

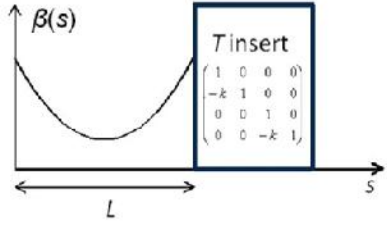

Figure 2: The ring outside of the nonlinear section has a transfer matrix of a thin axially symmetric lens [5].

Table 1. Main parameters of IOTA ring, setup with one nonlinear insert.

| | |
|---|---|
| Electron energy | 150 MeV |
| Number of bunches, particles per bunch | 1, $10^9$ |
| Ring circumference | 40 m |
| Synchrotron decay radiation damping time | 1 s |
| Equilibrium emittance, x & y, RMS | 0.04 mm-mrad |
| Betatron tunes, x & y | 5.3, 5.3 |
| Natural tune chromaticities, x & y | -11.4, -7.1 |
| Synchrotron tune | $5.3 \times 10^{-4}$ |
| RF harmonic number and voltage | 4, 1 kV |
| RMS energy spread and bunch length | $1.35 \times 10^{-4}$, 10.8 cm |

To approximate the inverse $^3$ dependence in Eq. (1), the 1.8-meter-long octupole channel consist of 18 identical octupoles with magnetic length of 10 cm. Figure 3 shows the distribution of their strength in the channel; $B_3$ is the normalized octupole gradient:

$$B_3 = \frac{1}{B\ldots} \frac{\partial^3 B}{\partial x^3} = 6\mathsf{r} \, \frac{1}{\mathsf{s}(s)^3} \tag{3}$$

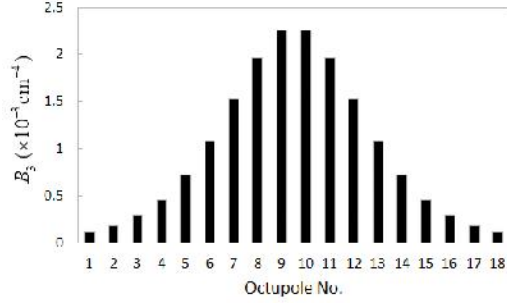

Figure 3: Octupole strength is maximal in the middle of the nonlinear section and rapidly decreases towards the ends; strength parameter = 110 cm$^{-1}$.

**Closer look at the approximate Hamiltonian**

Before proceeding to the numerical simulation let us estimate what kind of tune spread one can, in general, achieve in an octupole potential in an ideal case.

*Continuous octupole potential*

For the continuous octupole potential we obtain a Henon-Heiles-like system with a Hamiltonian, described by the Eq. (2). It has only one integral of motion – the Hamiltonian itself. The Hamiltonian (2) allows two integral manifolds $(x_N = P_x = 0$ and $y_N = P_y = 0)$, where the motion is one-dimensional. In this case the Hamiltonian reduces to:

$$H = \frac{1}{2}P_x^2 + \frac{1}{2}x_N^2 + \frac{1}{4}\Gamma x_N^4, \qquad (4)$$

where $x_N$ and $P_x$ are the normalized coordinates and momenta, and is the strength of the octupole nonlinearity. The Hamiltonian (4) is a constant of motion and therefore the system is integrable in this one-dimensional case. Because the potential of this system is a positive-definite form, the motion is stable and bounded for any initial amplitude. Therefore, the 1D system has an unlimited dynamic aperture and its frequency increases infinitely with amplitude (for > 0).

In general, apart from this very special case described above, the motion is in 2D. Due the lack of the second integral the motion the system has a finite dynamic aperture (Fig. 4a). The size of the region of stable motion is approximately defined by the distance to the unstable fixed points $P_x = P_y = 0, |x_N| = |y_N| = 1/\sqrt{2\Gamma}$. From our modeling we observed that a particle becomes unstable if its amplitude is greater than ~ 0.6/ $^{1/2}$. The region of regular motion is actually slightly smaller because of a thin layer of bounded chaotic motion near the boundary of the stable zone. The normalized acceptance of the system is

$$A = x_N^2 \sim (0.6r^{-1/2})^2 \sim 0.4/r \qquad (5)$$

Unlike the 1D case, the frequency shifts of a 2D system are both positive and negative. Figure 4 (b) depicts the spread of frequencies, obtained by tracking of $10^4$ particles with a Hamiltonian given by Eq. (2). The maximum positive tune shift is about 0.14, while the negative is about 0.26, resulting in the total frequency spread of 0.4, or, going back to the physical units, 40 % of the fractional tune.

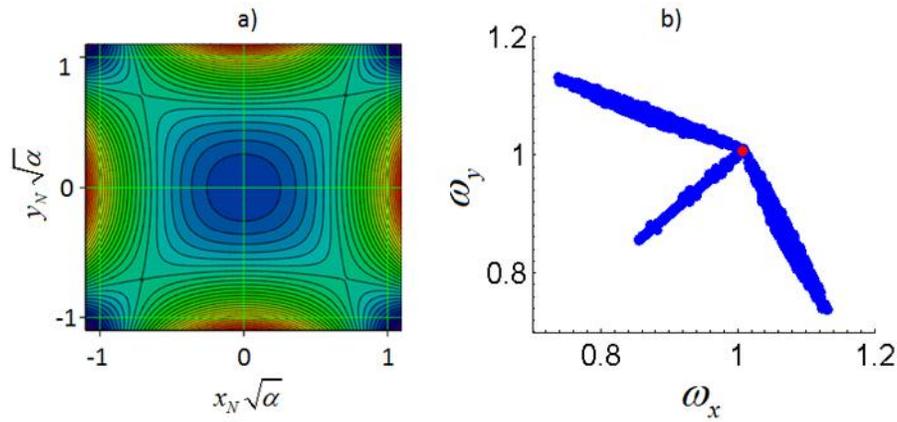

Figure 4: The motion remains bounded only for initial amplitudes $< 0.6^{-1/2}$, and the corresponding tune spread is 0.7. a) – Lines of equal potential of the Hamiltonian (2). b) – Frequency spread of the stable particles; the linear tune (without the octupole potential) is normalized to 1 (red dot).

*Practical implementation*

In practice, creating a continuous octupole potential is challenging, as it requires installing octupole magnets with properly scaled strengths throughout the ring. At IOTA we solve this problem by tuning the linear lattice of the ring outside the octupole channel to have a transfer matrix of a thin axis-symmetric lens (Fig. 2) and a phase advance of 2 n. Then in the normalized coordinates the transfer matrix is just an identity matrix **I**. A more traditional approach is to create the nonlinear tune spread with short octupoles, concentrated in one or several locations only.

A short octupole, added to an otherwise linear lattice, breaks down the integrability. In 1D, the motion is limited by $8^{th}$ order resonances $8\varepsilon = n$, thus the nonlinear shift of the tune cannot exceed

0.125. In 2D case, coupling resonances add to the 1D ones, significantly reducing the available phase space. The exact number for the maximum attainable tune spread is machine-specific and depends on the choice of bare betatron tunes (which are typically about equal). For example, a maximum tune shift of $5\times10^{-3}$ at 6 rms beam sizes, or 6 σ, can be obtained without damaging the dynamic aperture at the LHC [11].

One can achieve higher tune shifts by combining multiple octupoles into families and placing them in the locations of maxima and minima of beta-functions. A recent numerical study of SIS-100 synchrotron at GSI has reported the possibility of creating a 0.08 tune spread using 24 octupoles of two families distributed through the ring [12].

**Numerical model**

Because the analytical theory describes a considerably simplified Hamiltonian (Eq. (2)), we performed an accurate numerical tracking to check how it applies to a more realistic model of an accelerator. We simulated 3D particle dynamics in a realistic machine with betatron and synchrotron motion coupled through dispersion and chromaticity using Lifetrac particle tracking code [13] running on the Fermilab's Wilson cluster. The code was initially developed for beam-beam simulations but can also be used for 3D tracking of particles in the presence of an arbitrary nonlinear potential. Our model of the ring incorporated dipoles, quadrupoles, and an RF cavity – all ring elements, except the injection magnet, kickers, corrector magnets, and skew sextupoles, which are incorporated in the corrector magnets. It also included the parasitic nonlinear dipole fringe fields, which were simulated up to the 4$^{th}$ multipole order. The nonlinear potential (1) was created by a set of 10-cm-long hard-edge octupole magnets with their strengths governed by the expression (3).

The machine elements were represented by a number of thin kicks and drifts in a symplectic manner. The element slicing was performed in MAD-X [14] using the so-called Teapot algorithm [15]. After the slicing, the lattice data was imported from MAD-X into Lifetrac for the studies. Lifetrac implements 6-dimensional symplectic thins-lens tracking following [16]. It uses the paraxial approximation for the multipoles and properly treats non-paraxial effects in the drifts. This method proved to be accurate for the LHC and DAΦNE tracking studies [17].

We used two methods to analyze the tracking results: dynamic aperture plots and Frequency Map Analysis (FMA). The FMA, originally proposed by Laskar in [18], thanks to its high computing performance allows to rapidly scan through the phase space and determine frequency shifts as functions of betatron oscillation amplitudes, identify potentially dangerous resonances, and find the boundary of chaotic region [19]. A drawback of the FMA is that it finds only the single dominant frequency of particle motion and may fail to resolve multiple Fourier peaks, arising when a particle is close to a resonance. So in addition to the FMA we performed tracking for a time of synchrotron radiation decay(~ $10^6$ turns) to treat such situations and determine the dynamic aperture of the machine exactly. A smaller subset of initial amplitudes was used for this purpose. During this study the, initial amplitudes both in x and y were increased until a particle loss was observed.

**Simulation results and discussion**

In this section we, first, find the tune spread and dynamic aperture in an ideal IOTA lattice and compare it with the theoretical limits discussed above. Then we discuss some optimization issues regarding the phase advance in the nonlinear section and the strength of the nonlinearity. Then we study separately the effects of longitudinal dynamics, lattice errors, and magnet imperfections. Finally, we consider an application of the octupole channel for Landau damping.

*Ideal lattice*

First, let us find the maximum achievable tune spread for the case of an ideal IOTA lattice with one octupole section, no imperfections, and no initial synchrotron motion: $dE/E = 0$. We should note that although the particles are placed on-momentum, the synchrotron motion will still appear due to coupling between the horizontal and longitudinal degrees of freedom in our 3D model. The FMA footprints for this case are shown in the middle line in Fig. 5. We used $8192 = 2^{13}$ turns of tracking to create the FMA plots. The plot on the left shows the jitter of betatron frequencies as a function of normalized betatron amplitudes. The quantity plotted is $\log\sqrt{\Delta Q_x^2 + \Delta Q_y^2}$, where $\Delta Q_x$ and $\Delta Q_y$ are the variations of the horizontal and the vertical tunes during the tracking interval. A wide resonance line at 45 degrees in the amplitude subplot is the difference resonance $Q_x - Q_y = 0$. Tracking shows that particles in this region remain stable due to the absence of crossings with other resonances, which could lead to transfer of energy between degrees of freedom and eventually particle loss. The dynamical aperture for this case is about $0.5/\sqrt{\Gamma}$. The phase space footprint corresponding to the stable region is depicted on the right. Tune shifts in $Q_x$ and $Q_y$ range from -0.05 to +0.03, creating a total maximum tune spread of 0.08 for a beam occupying the entire stable region.

FMA also shows numerous thin resonance lines (green). Those are high order resonances (typical order in the range 10 - 20) involving both betatron and synchrotron motion. They are unlikely to be excited in the real machine and consequently would not disturb the particle motion. In our tracking studies the particles stay stable when placed in these regions for $10^6$ turns, a period which is comparable with the synchrotron damping time.

One can compare the performance of the proposed optics with the idealized case of 2D continuous octupole in the normalized variables of maximum amplitude $x_N \sqrt{\Gamma}$ and frequency spread $\Delta Q$ (Fig. 5, top). The resulting dynamic aperture and maximum frequency shift are within 20-30% of the theoretical limits (Table 2). A similar comparison can be done with the performance of a single short octupole installed in the ring. The octupole was placed in the middle of the nonlinear section and its strength was equal to the integrated strength of the octupole channel. The short octupole yields a smaller dynamic aperture than an 18-octupole channel (Fig. 5, bottom). The total tune spread of the channel is about two times greater than that of the short octupole (Table 2). Note that the tune spread achievable with a short octupole might be an overestimate as our model did not include some ring elements like sextupoles. In general, the tune spread for this case depends on the widths of the resonance lines and the position of the working point relative to them.

Even in an ideal ring with no errors the Hamiltonian does not remain exactly constant because of the presence of the synchrotron motion, which is not accounted for in the Eq. (2). Despite the presence of synchrotron motion the Hamiltonian (2) is conserved on a level of $10^{-4}$-$10^{-3}$ within the ring's dynamical aperture of $0.5/\sqrt{\Gamma}$ (Fig. 6, left). In the absence of the synchrotron oscillations (particles are placed on-momentum, and the ring is represented by a transfer matrix with no coupling between the transverse and the longitudinal degrees of freedom) the deviation of the Hamiltonian from its mean value is at least an order of magnitude smaller. In this case, the Hamiltonian is conserved on a level $10^{-5}$-$10^{-4}$ up to the amplitude of $0.5/\sqrt{\Gamma}$. At the larger amplitudes it is conserved at the level of $10^{-3}$ for nearly the whole area of stable motion (Fig. 6, right). This remaining variation of the Hamiltonian seems to be caused by the approximation of the continuous potential in (2) by a set of discrete magnets.

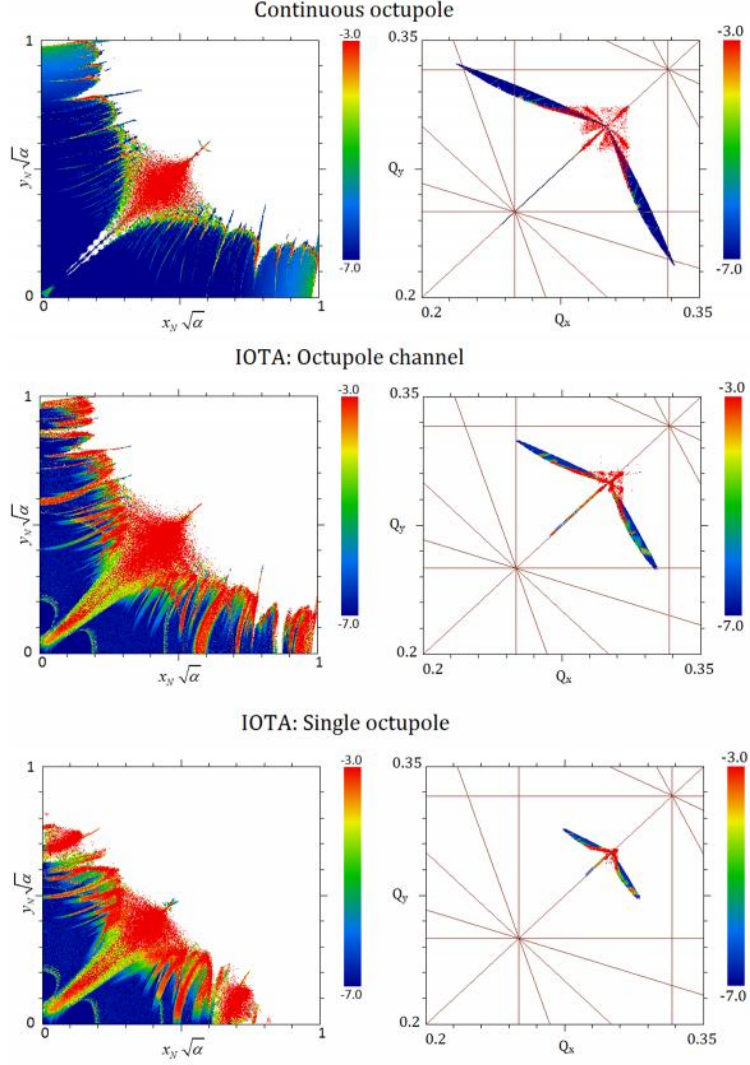

Figure 5: FMA footprints of phase space in average betatron amplitudes (left) and betatron frequencies (right). Thin brown lines show resonances up to the 4$^{th}$ order. Octupole strength parameter = 400 cm$^{-1}$.

Table 2. Performance comparison

| Dynamic system | Maximum $x_N$ | Frequency spread |
| --- | --- | --- |
| Continuous octupole (Eq. 2) | $0.6/\sqrt{\Gamma}$ | 0.12 |
| IOTA: Octupole channel | $0.5/\sqrt{\Gamma}$ | 0.08 |
| IOTA: Single octupole | $0.4/\sqrt{\Gamma}$ | 0.04 |

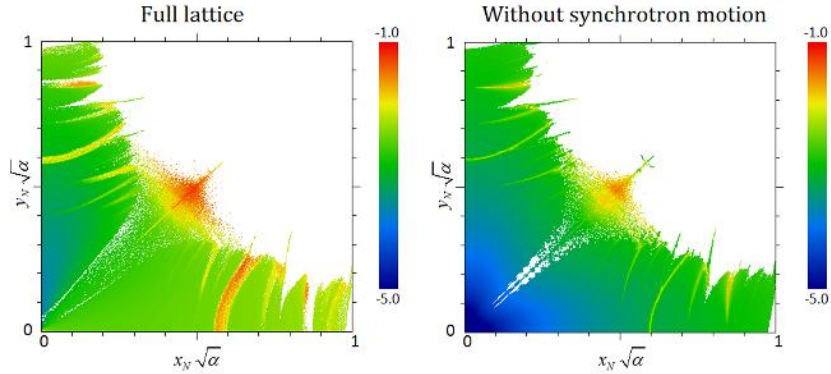

Figure 6: Relative standard deviation of the Hamiltonian form its mean value is of the order of or below $10^{-3}$ for most amplitudes of stable betatron oscillations (one octupole channel, strength parameter  = 400 cm$^{-1}$). Left – full IOTA lattice with synchrotron motion, coupled to betatron motion through dispersion and chromaticity; right – no synchrotron motion, the particles are places on-momentum.

*Some optimization considerations*

Figure 7 shows the dependence of the maximum spread of betatron frequencies and the maximum amplitude of stable motion on the strength of the octupole potential. For small strength parameters  , the amplitude is limited by the physical aperture and the maximum tune shift increases with  . For sufficiently large  , the dynamical aperture is limited by resonances (Fig. 5), and consequently the maximum tune shift is independent of the potential strength. Thus there is a value of  , which yields the greatest dynamical aperture for the maximum possible tune shift. Our modelling shows that in IOTA this value is  = 100 cm$^{-1}$. The dynamic aperture for this optimal strength is 20$\sigma$ of the electron beam and the maximum tune spread is 0.08.

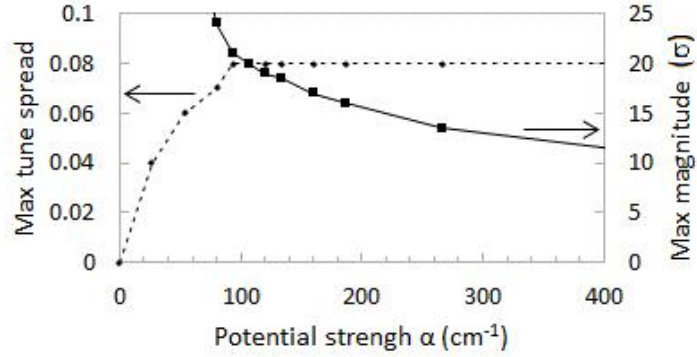

Figure 7: Maximum tune spread is achieved at the strength of octupole potential $= 100$ cm$^{-1}$; further increase of nonlinearity leads only to reduction of dynamical aperture.

Since the octupole focusing strength is quadratic in betatron amplitude and the maximum octupole tune shift is independent of the strength, above a certain value, there is a tradeoff between the dynamical aperture and the rms tune shift:

$$dQ_{RMS} = \frac{dQ_{max}}{A_{max}^2} \qquad (6)$$

Figure 8 depicts a dependence of maximum achievable spread of betatron tunes on the phase advance over the drift section. For the purpose of this simulation the rest of the ring was treated as an axially-symmetric focusing transfer matrix with an integer phase advance. One can see that the dependence is close to linear and can be approximated as

$$dQ_{max} \sim 0.4\Delta\Psi, \qquad (7)$$

where $\Delta\Psi$ stands for the phase advance in the drift. The 40% slope of the dependence agrees with the theoretical estimate made in the approximation of the continuous octupole potential (Fig. 4). The deviations from the straight line can be explained by the presence of low order resonances in the ring. Its tunes change with $\Delta\Psi$, since the phase advance in the ring outside of the nonlinear section should always be a multiple of 2 . When the tunes get close to a low order resonance it starts to limit the dynamic aperture and the maximum tune spread decreases (point 0.35 on the plot).

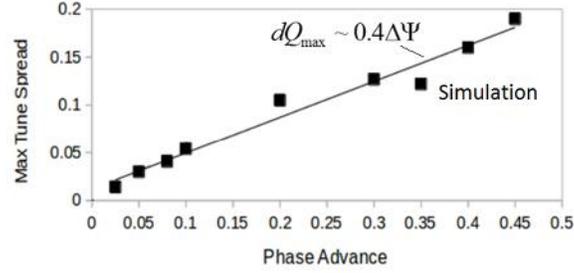

Figure 8: Maximum tune spread increases linearly with the phase advance in the insert. One insert in the ring, the rest of the ring was modelled by a linear transfer matrix.

In practice, it is hard to create a phase advance close to 0.5 over a straight section, since that would require a creation of a very low β-function in the middle of the drift. The 0.3 phase advance in the current IOTA design allows keeping the β-functions at relatively moderate values: 0.65 m in the middle of the 1.8-m-long section and 1.9 m on the ends, while still providing a high frequency spread. If a greater nonlinear tune spread is needed, one can increase the number of nonlinear sections in the ring. The symmetry of the IOTA lattice allows for two nonlinear inserts.

*Effect of synchrotron motion*

In the ideal case described above, the particles started with no initial synchrotron motion: $dE/E = 0$. The motion arose due to coupling between the horizontal and longitudinal degrees of freedom in our 3D model. The average amplitude of synchrotron oscillations increased with betatron amplitude up to $3\sigma_s$ at the boundary of the stable region.

In a non-ideal case, for the particles with initial synchrotron oscillation amplitudes $A_s < 1\sigma_s$, the resulting frequency maps differ from the ideal picture insignificantly. In those cases the initial synchrotron amplitude is much smaller than the amplitude, arising from the betatron motion due to betatron-synchrotron coupling. However, for $A_s > 6\sigma_s$ the synchrotron motion starts to destroy the dynamics significantly, and the dynamical aperture can be reduced by 15% or more.

Apart from the synchrotron motion, another factor affecting the integrability is the chromaticity of betatron tunes. It introduces a first order perturbation to the Hamiltonian if x- and y-chromaticities are not equal [20]. To obtain numerical estimates of this effect a part of the ring was replaced by a transfer matrix with phase advances in *x* and *y* depending on $p_z$. By changing this artificially introduced chromaticity we were able to get a dependence of dynamic aperture on the difference of betatron frequencies $\Delta Q_{xy} = (\xi_x - \xi_y) \cdot dp/p$, depicted in Fig. 9. One can also see that the dynamical aperture is the greatest when $\xi_x \approx \xi_y$, as predicted in [20]. The natural chromaticities of IOTA: -11, -7 are close to this condition. Dynamical aperture decreases as the difference in tunes, created by unequal chromaticities, shifts further away from 0. For $|\Delta Q_{xy}| > 10^{-3}$ it is less than ½ of the maximum value.

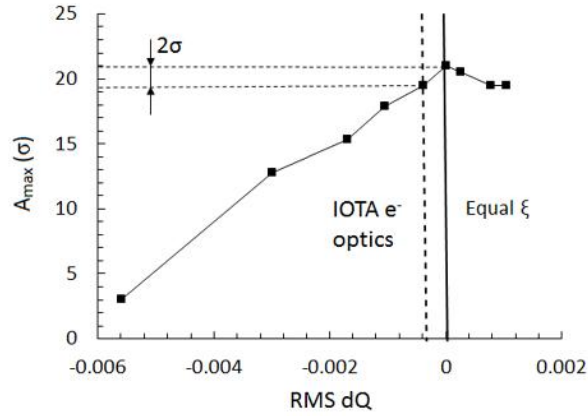

Figure 9: Tune mismatch, created by chromaticity, leads to reduction of dynamic aperture. For electron operation the decrease is about 2 rms beam sizes. = 100 cm$^{-1}$.

Since most of the particles in IOTA electron bunch will have $dp/p$ less than or of the order of $(dp/p)_{RMS}$, the longitudinal motion does not affect the transverse dynamics significantly. Though it might become an issue for the proton experiment at IOTA because of the larger momentum spread of the proton beam (~ 10$^{-3}$).

*Tolerance to design errors and magnet imperfections*

To study the influence of errors in the linear lattice, the strengths of several quadrupole magnets were manually varied to detune $Q_y$ from its design value 0.3, while leaving $Q_x$ unperturbed. Previous similar studies of fully integrable potential show that it requires a precise matching of beta functions and tune to the design values. An error of more than 10$^{-3}$ in phase advance over the ring breaks the integrability [9]. In contrast to the fully integrable case, the octupole channel optics is more robust against errors in linear focusing: errors of the order of 10$^{-3}$ in betatron tunes produce no visible effect on particle dynamics, $\Delta Q_y = 10^{-2}$ ($\Delta s / s$ in the middle of the insert of 0.06) leads to a reduction of dynamic aperture of ~ 10%, and for $\Delta Q_y = 10^{-1}$ ($\Delta s / s = 0.2$) dynamic aperture shrinks by a half (Fig. 10).

To estimate the influence of imperfections in the nonlinear potential on particle dynamics, a normally distributed error was added to each of the octupole magnets. The simulations show that for a 10% RMS error in octupole strength, the reduction of dynamical aperture is negligible.

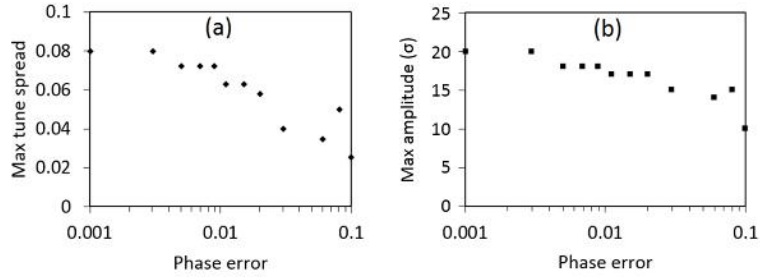

Figure 10: Detuning of linear lattice leads to decrease of maximum tune spread (a) and dynamic aperture (b). Octupole strength = 100 cm$^{-1}$

*Octupole channel for Landau damping*

The tune spread, created by the octupole channel, could be used to enhance Landau damping of collective instabilities in high intensity beams. Note that there are other ways to create the spread of betatron tunes: for example, tune chromaticity. But linear chromaticity provides Landau damping only in the cases a coasting beam or very long bunches, where the period of synchrotron oscillations is significantly greater than the characteristic growth time of the instability: $T_s \gg \tau_{inst}$. In a more common case of short bunches with relatively low $T_s$ the betatron tune shift due to chromaticity, averaged over the synchrotron period, is zero to the first order: $\langle \langle dp/p \rangle_s \rangle = 0$ for every particle in the bunch, and thus there is no damping. Although it does not provide damping (at the first order), tune chromaticity affects the instability through its growth rate. Correcting the chromaticity reduces the growth rate or even stabilizes the most unstable collective mode, which is typically the lowest frequency one [21]. The amount of correction one can achieve without significant damage to dynamic aperture and beam lifetime is specific for every accelerator as it depends on the working point, beam intensity, etc. In some cases, it may be more beneficial to create additional Landau damping by introducing nonlinear tune spread with octupoles as described in the present paper.

Knowing the tune spread one can estimate the strength of Landau damping. According to Burov [21], the damping rate of space charge driven collective instabilities is quadraticaly proportional to RMS nonlinear tune spread:

$$\Lambda_x = \frac{\Delta Q_{xx}^2}{\Delta Q_{SC}} F(a_{xx}, a_{xy}),$$
$$\Delta Q_x = a_{xx} J_x + a_{xy} J_y = \Delta Q_{xx} + \Delta Q_{xy}$$
(8)

where $\Delta Q_{SC}$ is a space charge tune shift and $F$ is a damping factor. For the proposed parameters of PIP-II Recycler upgrade [22], a single octupole insert is capable of creating the tune spread up to $5*10^{-3}$, but the downside is a relatively small resulting dynamic aperture: about 4 sigma in ideal case with no imperfections. A possible solution for achieving a greater spread may be using more than one octupole insert. We shall also notice that [21] gives only a qualitative estimate of the damping rate, and a detailed numerical study of Landau damping in the presence of nonlinear tune spread is required. Such research is currently under way at FNAL, and the first simulation results are consistent with the theory [23].

**Conclusion**

We have studied single particle dynamics for the Henon-Heiles-like nonlinear optics in the Integrable Optics Test Accelerator using a realistic 3D model of the ring. The nonlinear potential was created by a set of octupole magnets, placed in a special manner to approximate an autonomous 2D Hamiltonian system. Although this system is non-integrable by design, the presence of an approximate invariant of motion seems to help achieving a large nonlinear spread of betatron tunes.

The numerical simulation allows to determine nonlinear tune shifts and dynamical apertures in the presence of lattice errors and magnet imperfections. The results show that a tune spread of 0.08 can be achieved with one channel of conventional octupole magnets, installed in the ring. This spread is significantly larger than the ~$5*10^{-3}$ spread typically obtained with octupoles in accelerators. Despite the strong nonlinearity the ring retains a relatively large dynamic aperture of 20 . This aperture is about 80% of the theoretical limit for a continuous octupole potential.

The effect of longitudinal motion and lattice chromaticity on the dynamics of electrons in the ring are insignificant, although they may create certain challenges for proton operation due to higher momentum spread. The study has also shown that the octupole optics is robust to lattice errors and imperfections. Errors of up to 0.01 in betatron tunes, 6% in beta-functions, and 10% in the potential itself lead to less than 15% reduction of dynamical aperture. This level of precision can be realistically achieved in future high-intensity circular accelerators.

Future circular accelerators can benefit from employing the described nonlinear optics by creating significant tune spreads to enhance Landau damping of collective beam instabilities. Because all the nonlinearity is created locally, in the octupole channel, the approach might also be useful for upgrading existing machines.

The proof-of-principle experiment is currently under construction at the IOTA facility at Fermilab. A prototype octupole magnet has been assembled and tested and the assembly of the whole 18-magnet octupole channel is scheduled for the 2017.

**Acknowledgements**

The authors are grateful to Alexei Burov (FNAL) for numerous useful discussions, and to Dmitri Shatilov (BINP) for his help with the tracking code.